# Exploiting and engineering the optical field-enhancement in metal nanoparticle based circuits

M. A. Mangold, C. Weiss, B. Dirks, and A. W. Holleitner[*]

*Walter Schottky Institut and Physik Department, Technische Universität München, Am*

*Coulombwall 3, D-85748 Garching, Germany.*

*\*Corresponding author: holleitner@wsi.tum.de*

We study the impact of optical field-enhancement effects on the optoelectronic properties of metal nanoparticle arrays. Applying a focused ion beam lithography in combination with an electron beam deposition technique we can pattern electrical contacts in a way to give rise to an electromagnetic field-enhancement. A further field-enhancement is achieved by the plasmon resonance of the particles. Both effects are directly observed in the photoconductance of the arrays.

Arrays of metal or semiconducting nanoparticles (NPs) have outstanding optical and optoelectronic properties, which can be fine-tuned by both the physical properties of the individual NPs and the overall response of the arrays.[1-3] Recent interest in semiconducting NP devices is spurred by the possibility to fabricate photodetectors exploiting a multiexciton generation in NPs.[4] In metal NP arrays, the excitation of surface plasmons can lead to a strong electromagnetic field enhancement at the surface of NPs at visible and near-infrared wavelengths.[5] Hence, recent work has focused on surface enhanced Raman spectroscopy[6-8] as well as on the photonic band gap[9] and photoelectrochemical properties of NP arrays.[10] In the context of molecular electronics, densely packed two-dimensional (2D) metal NP arrays have been used to electrically contact molecules.[11,12] Optical studies have revealed a strong plasmonic absorption in such arrays, where alkanes or conjugated molecules were inserted.[13,14] The plasmonic field-enhancement gives rise to promising applications as in solar cells,[15,16] photodetectors,[17-19] bio-sensors,[20,21] and light-emitting diodes.[22]

Here, we demonstrate how to exploit and engineer the electromagnetic field-enhancement in well ordered, electrically contacted 2D arrays consisting of metal NPs coated with alkane thiols. On the one hand, a field-enhancement is achieved by the arrangement of patterned electronic contacts. On the other hand, the field is enhanced by a surface plasmon resonance in the NPs. For the first effect, we introduce a focused ion beam (FIB) lithography in combination with an electron beam induced deposition (EBID) technique. The combination enables us to electrically contact NP arrays in different geometries. We observe an increase of the field-enhancement with a smaller distance between the electronic contacts, directly resulting in an increased optoelectronic response of the NP arrays of 20-30 %. We find good agreement between the data and a finite-difference time-domain (FDTD) simulation. This extrinsic field-enhancement effect may prove useful to optimize the optoelectronic response of circuits based on both semiconducting and metal NP arrays. We verify the plasmonic enhancement by photon energy dependent measurements. Again, the experimental findings are in good agreement with FDTD-simulations. Our results may help to design submicron optoelectronic sensors with functionalized NPs and molecules embedded.[23,24] In particular, the intrinsic field-enhancement due to plasmons occurs at interparticle locations, where conducting molecules can be incorporated.[11,12]

Starting point are 2D arrays of alkane coated gold NPs, fabricated as in [11,13]. The NPs are ordered in 2D hexagonal arrays with a lattice constant of ~12 nm and a NP-NP distance of 2-3 nm. The arrays initially have dimensions of 20 × 8 μm$^2$, and they are electronically contacted by gold electrodes. By the help of FIB milling and EBID, the size of the arrays is then reduced to minimum dimensions of ~100 ×100 nm$^2$.[25] Fig. 1(a) depicts a sketch of the resulting device with a spatially trimmed NP array in the center. In a first step, the width of the NP arrays is reduced by FIB milling [gray areas in Fig. 1(a)]. Then, the gold electrodes are prolonged by an EBID of platinum (~30 %) and carbon (~70 %).[26] Hereby, the arrays are electronically contacted by the EBID contacts [dashed areas in Fig. 1(a)]. Fig. 1(b) presents a scanning electron microscope (SEM) image of such a spatially engineered array with dimensions of ~400 × 400 nm$^2$. The gold clusters adjacent to the NP array, which are remnants of the FIB milling, do not participate in the electric conduction of the device. Fig. 1(c) shows a high resolution SEM image of such an array. Importantly, the hexagonal ordering is not disturbed by the FIB and EBID. On top, the spatially trimmed arrays still exhibit a linear I-$V_{SD}$ characteristic with a conductance of nS for a source-drain voltage |$V_{SD}$| ≤ 1V (data not shown). This is consistent with the results of macroscopic arrays.[11,13]

All measurements are performed at room temperature at a pressure of ~$10^{-6}$ mbar. Optical excitation occurs by focusing the light of a mode-locked titanium:sapphire laser with a repetition rate of 76 MHz through a microscope objective onto the arrays. By exploiting a photonic glass fiber, laser light is available in the spectral range of 520 nm < $\lambda$ < 1000 nm. With a spot diameter of ~2 µm the light intensity $I_{OPT}$ is about 1 kW/cm$^2$ for all $\lambda$. For the optoelectronic measurements, we chop the laser at frequency $f_{CHOP}$. The resulting current is amplified by an I-V converter. The optically induced current $I_{PHOTO}$ is detected with a lock-in amplifier utilizing the reference signal provided by the chopper.[13] We find that $I_{PHOTO}$ depends linearly on $V_{SD}$ and we detect no finite value of $I_{PHOTO}$ at zero bias [Fig. 1(d)]. Thus, $I_{PHOTO}$ is due to an optically induced change of conductance, and we use the term "photoconductance", i.e. $G_{PHOTO} = I_{PHOTO}/V_{SD}$, to describe the phenomena.

We measure $G_{PHOTO}$ and concurrently the reflection of the laser spot, while the position of the sample is moved in the x-y-plane. In Fig. 1(e), the spatially resolved $G_{PHOTO}$ is superposed to the reflection image of a NP array. We clearly observe a photoconductance signal above noise level only at the position of the spatially trimmed array [center in Fig. 1(e)]. We point out that there is no signal at the position of the EBID contacts or the FIB milling. Fig. 1(f) shows the spatially resolved $G_{PHOTO}$ taken in steps of 200 nm. The spatial full width half maximum (FWHM) of $G_{PHOTO}$ matches the laser spot size (~2 µm).

First, we consider the external enhancement effects, i.e. the field-enhancement caused by the EBID contacts. To this end, we perform FDTD simulations of the distribution of a linear polarized electromagnetic field irradiated onto two 1 µm wide electrodes with a gap of 200 nm. For the contacts, the optical properties of carbon are assumed, since the EBID material consists to 70 % of carbon. In Fig. 2(a), the light polarization is chosen such that the electrical field is oriented perpendicular to the electrodes (arrow). For a parallel orientation [arrow in Fig. 2(b)], we calculate a field-enhancement of ~20 % compared to the perpendicular one. Importantly, this effect enhances the optoelectronic response of the NP arrays. Fig. 2(c) depicts $G_{PHOTO}$ of a NP array as a function of the rotation angle of the light polarization (black squares). $G_{PHOTO}$ oscillates with a period of 180°. The line represents a sinusoidal fit to the data, and the inset depicts a fast Fourier-transformation (FTT). The relative amplitude of the oscillation, i.e. the amplitude of the oscillation divided by the mean $G_{PHOTO}$, for a 200 nm gap is (23 ± 6) %; in good agreement with the FDTD-simulations. Further experimental evidence for the oscillation being caused by an antenna effect of the EBID contacts is given by the decrease of the oscillation with increasing electrode distance. Fig. 2(d) shows the relative oscillation amplitude against the contacts' distance (black squares). The amplitude decreases gradually with increasing distance up to ~3 μm. The line indicates the 2 μm wide, Gaussian shaped laser spot. For an electrode gap larger than ~3 μm, the laser does not illuminate the electrodes, and in turn, the field-enhancement effect vanishes. We explain the remaining oscillation of ~5% by a polarization dependent absorption of the optical set-up.

Now, we examine the intrinsic field-enhancement effects, i.e. the field-enhancement caused by the hexagonal ordering of the NPs in a 2D array. The frame in Fig. 3(a) indicates the unit cell which we use in the FDTD simulations with periodic boundary conditions. Figs. 3(b) and (c) show the electrical field distribution for orthogonal light polarizations in 2D plots. In each figure, two unit cells are concatenated to accentuate the field strength between two NPs. In Fig. 3(b) the electric field vector of the light is aligned parallel to an axis connecting two neighboring NPs (arrow). The largest field amplitudes are located in between two adjacent NPs. Turning the electric field orientation by 90° [arrow in Fig. 3(c)], the field-enhancement is smaller by ~20 %. We argue that the intrinsic field-enhancement is caused by a collective plasmonic excitation of the NPs. This is corroborated by a simulation of the wavelength dependence of the field-enhancement. Fig. 3(d) shows the intensity of the electrical field at the position between two NPs as a function of the photon wavelength. We find a maximum of the field-enhancement at ~600 nm. This value coincides with the surface plasmon resonance of arrays made from dodecane thiol coated NPs as predicted by the Maxwell-Garnett effective medium theory.[13,14,27] We observe the surface plasmon resonance also in the photoconductance of the arrays. Fig. 3(e) shows $G_{PHOTO}$ of an array with dimensions of ~390 × 600 nm$^2$ as a function of wavelength. The maximum at ~600 nm is in accordance with the simulations. We note that the maximum can clearly be detected despite the spatial fluctuations as apparent in Fig. 1(f), and that in macroscopic NP arrays, the maxima can be detected with a lessened noise level.[13]

To conclude, we exploit FIB and EBID lithography to examine the impacts of both an external antenna effect caused by the electrodes connecting the NPs and an internal field-enhancement caused by the hexagonal ordering of NPs in 2D arrays. We find that both effects lead to an enhancement of the photoconductance of NP arrays. The spatial engineering of optoelectronic circuits may prove useful for the design of nanoscale sensors, where for instance molecules are embedded into the gaps between two adjacent NPs.

We thank M. Calame and Ch. Schönenberger for fruitful discussions and the German excellence initiative via the "Nanosystems Initiative Munich" (NIM) and the DFG SPP 1391 for financial support.

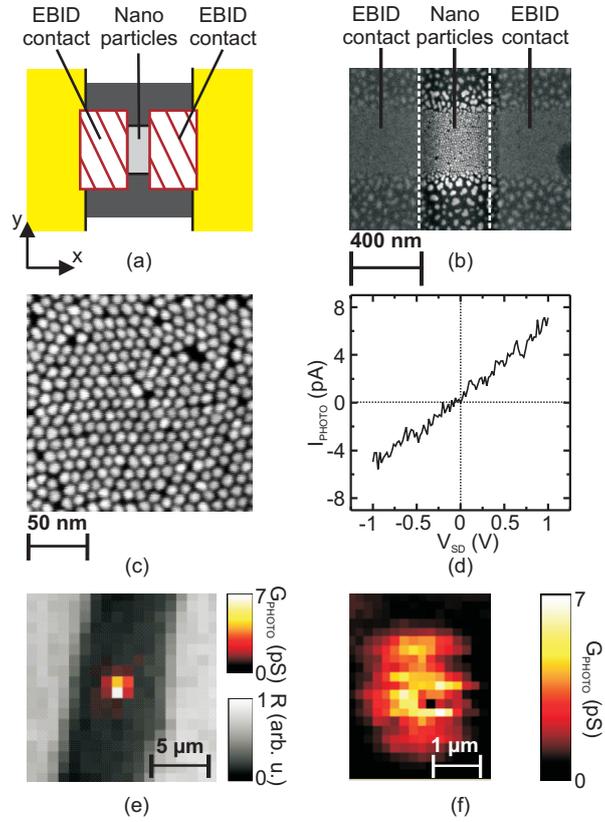

Figure 1

**Fig. 1.** (color online) (a) Sketch of a spatially trimmed nanoparticle (NP) array with contacts patterned by electron beam induced deposition (EBID). (b) Scanning electron microscope (SEM) image of a gold NP array with EBID contacts. (c) High-resolution SEM image of such a NP array. (d) $I_{PHOTO}$ vs. $V_{SD}$ characteristic of a NP array ($\lambda$ = 600 nm, $I_{OPT}$ = 0.7 kW/cm$^2$, $f_{CHOP}$ = 873 Hz). (e) Superposition of the photoconductance (center) and reflection (peripheral signal) as a function of laser spot position of the array from Fig. 1(b). (f) High resolution scan of the photoconductance as a function of laser spot position ($\lambda$ = 600 nm, $V_{SD}$ = 0.2 V, $I_{OPT}$ = 1 kW/cm$^2$, $f_{CHOP}$ = 117 Hz in (e) and (f)).

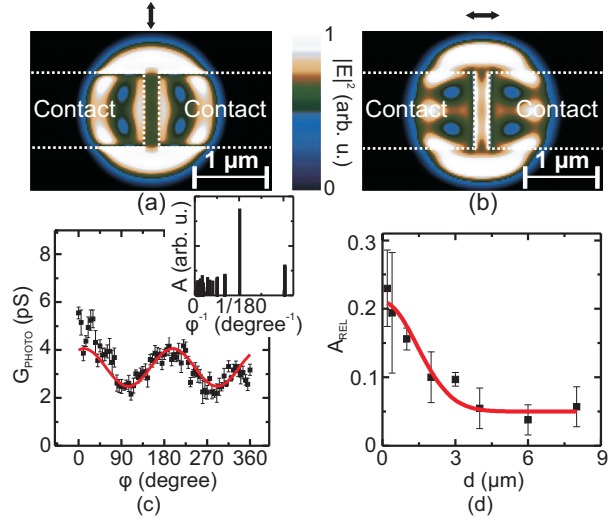

Figure 2

**Fig. 2.** (color online) (a) and (b) Simulations of the electrical field distribution induced by a 2 μm wide laser beam at λ = 600 nm focused on 1 μm wide and 20 nm thick contacts with a distance of 200 nm. Arrows denote the orientation of the electric field polarization. (c) Photoconductance of an array as a function of the rotation angle of the polarization (black squares, λ = 600 nm, $V_{SD}$ = 0.2 V, $I_{OPT}$ = 2.3 kW/cm$^2$, $f_{CHOP}$ = 3378 Hz). Line represents sinusoidal fit. Inset: fast Fourier transformation (FFT) of the data. (d) Relative amplitude of the sinusoidal fit as in (c) as a function of the distance between the EBID contacts. Line indicates the 2 μm wide Gaussian shape of the laser spot.

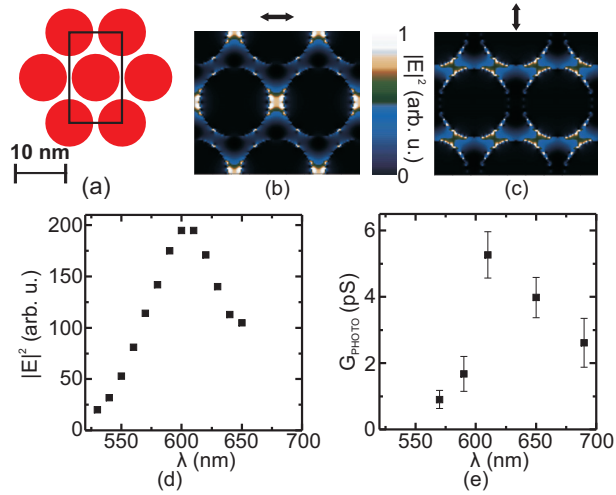

Figure 3

**Fig. 3.** (color online) (a) Sketch of a hexagonally ordered NP array. Frame indicates the unit cell of the finite-difference time-domain (FDTD) simulation. (b) and (c) 2D plots of the electric field intensity in an NP array with NP diameter of 10 nm and a 2 nm distance between adjacent NPs embedded in an alkane matrix. In (c) the polarization is turned by 90 degree with respect to (b) (see arrows). (d) Maximum electrical field strength as a function of $\lambda$ found in simulations with polarization as in (c). (e) Photoconductance of a NP array as a function of $\lambda$ ($V_{SD}$ = 2 V, $I_{OPT}$ = 0.3 kW/cm$^2$, $f_{CHOP}$ = 873 Hz).